\documentstyle[sprocl,epsf]{article}
\def\ifmath#1{\relax\ifmmode #1\else $#1$\fi}%
\def\rd{\ifmath{{\mathrm{d}}}}
\def\rt{\ifmath{{\mathrm{t}}}}

\def\tot{\ifmath{{\mathrm{tot}}}}
\def\inv{\ifmath{{\mathrm{inv}}}}
\def\long{\ifmath{{\mathrm{long}}}}
\def\side{\ifmath{{\mathrm{side}}}}
\def\out{\ifmath{{\mathrm{out}}}}

\def\ov{\over\displaystyle\strut}
\def\dst{\displaystyle\strut}

\begin{document}
\bibliographystyle{unsrt}    % for BibTeX - sorted numerical labels by order of
                             % first citation.
\null
\vskip - 1.5 truecm
\rightline{CU-TP-797 (1996)}
\vskip 1.0 truecm
\title{DATABASE ON THE NET FOR PARTICLE CORRELATIONS AND SPECTRA}
\author{J. SCHMIDT-S\O RENSEN}
\address{Physics Institute, University of Lund,
Professorsgatan 1, S -- 223 63 Lund, Sweden\\
janus@quark.lu.se}

\author{T. CS\"ORG\H O}
\address{Department of Physics, Columbia University,
538 West 120th, New York, NY 10027\\
and MTA KFKI RMKI, H -- 1525 Budapest 114,
POB 49, Hungary\\
csorgo@nt1.phys.columbia.edu~,  csorgo@sunserv.kfki.hu}

\maketitle\abstracts{
A fully interactive database is presented, which provides user-friendly
 access to experimental results on single-particle spectra and 
two-particle correlations (particle interferometry or HBT-effect). 
The database is available on the World-Wide-Web at: 
{\tt http://www.quark.lu.se/$\tilde{\,\,\,\,}$janus/hbt
\underline{ }home.html}~.}

\section{Idea of and Motivation for the Database}

A fully interactive database could make the access to the detailed
experimental data more efficient than  just looking up the data
(figures, functions or tables) in scientific journals. We report here 
about the creation of such a database, where the experimentally 
determined parameters can be archived, together with the experimentally 
measured distributions, using the framework of the presented database. 
Additional information including pictures, links to related papers, PAW 
kumac files etc. could be stored together in such a searchable database. 
It is based on a user-friendly environment  and is easily accessible through 
the Internet.

In general, the structure of the database is quite generic
and its potential is in no way restricted to a certain type of data.
In particular, we started the creation of this database in order to 
prepare a catalyst  for a more detailed communication between experimentalists
and theoreticians working in the field of particle correlations and spectra.

\section{Definitions}

There has been recently a substantial increase in the number of publications
related to the study of two-particle correlations in the field of high-energy
heavy-ion physics. Short-range correlations of various pairs of elementary 
particles yield information on the space-time extent of the region which 
produces particles with a given momentum in high-energy collisions. Recently, 
it became clear that a simultaneous analysis of two-particle correlations 
and single-particle spectra may yield some clues about the full size of the 
interaction region (core), about the flow, the density and the temperature 
profiles.\cite{nr,1d,3d,halo,uli_s}  Further, the two-particle correlation 
function has been experimentally analyzed in  a more and more sophisticated 
manner.\cite{bengt,zajc,qm96,3d,urev} The two-particle correlation function 
is defined as
\begin{equation}
	C({\bf\Delta k},{\bf K})  =  \frac{\langle n \rangle^2}
				{\langle n(n-1) \rangle} \,
				\frac{ N_2({\bf p}_1,{\bf p}_2) }
				{N_1({\bf p}_1) N_1({\bf p_2})}\ ,
\end{equation}
where $\langle n \rangle$ is the mean multiplicity, 
$\langle n(n-1) \rangle$ is the second factorial moment of the
multiplicity distribution, $N_2({\bf p}_1,{\bf p}_2)$ is the two-particle
inclusive invariant momentum distribution, $N_1({\bf p}_1)$ is the 
single-particle inclusive invariant momentum distribution (IMD) or 
single-particle spectrum, ${\bf p}_1$ stands for the three-momentum of 
particle 1, the relative momentum is denoted by 
${\bf \Delta k} = {\bf p}_1 - {\bf p}_2$ and the mean momentum is
denoted by ${\bf K} = 0.5 ({\bf p}_1 + {\bf p}_2)$.

The invariant momentum distribution (IMD) for a 
given type of particle stands for the (single - or two-particle)
inclusive invariant momentum distribution, 
\begin{eqnarray}
   N_1({\bf p}) & = & E \frac{\rd N }{\rd{\bf p}}\, = \, {\dst E \over 
\sigma_{\tot}} {\dst \rd \sigma \ov \rd{\bf p}}, \\ 
   N_2({\bf p}_1,{\bf p}_2) & = & E_1 E_2 
	\frac{\rd N }{\rd{\bf p}_1 \rd{\bf p}_2} \, = \, 
	 {\dst E_1 E_2 \over \sigma_{\tot}} {\dst \rd \sigma 
         \ov \rd{\bf p}_1 \,  \rd{\bf p}_2}, 
\end{eqnarray}
where $\sigma_{\tot}$ is the total inelastic cross-section. It is assumed 
that particles 1 and 2 are identified and some kind of effective final-state 
interaction creates short-range correlations between them. For identical 
particles, such short-range correlations may arise due to Bose-Einstein or
Fermi-Dirac statistics, for charged particles due to the Coulomb
interaction, and for strongly interacting particles due to strong final 
state interactions. Most frequently correlations between identical particles 
are studied. Recently, a systematic study of short-range correlations was 
reported by W.A. Zajc at the HBT'96 conference,\cite{ztr} which investigated 
the correlations between pions, kaons and protons in any
pair-combination, including particles and anti-particles.  

The two-particle correlation function depends in general on 6 independent
momentum components. Currently, advanced measurements attempt to determine
this correlation as a function of 5 variables, assuming azimuthal  
symmetry.\cite{qm96} Preliminary data are becoming available for 6 dimensional
correlation functions, too~\cite{misk}. A correlation function of order 
$j$ depends in general on $3 \times j$ independent variables. These data 
are then analyzed in terms of simple parameterizations and only the fitted 
parameters are published in a number of cases. However, the measured 
higher-order distributions are frequently not available for the scientific 
community. E.g., very few two- or three-dimensional two-particle correlation 
functions are published in the literature, although a lot of 
parameterizations are published for these reactions. Similarly, the present 
practice of publishing measured single-particle spectra relies frequently 
on the publication of the slope parameters of the invariant momentum 
distribution, or on the measurement of the rapidity distribution.
Recent theoretical developments and also recent measurements suggest
that the spectrum is not factorizable, the rapidity and the transverse-mass 
dependence is coupled. Thus, at least the two-dimensional 
$\frac{\displaystyle \rd n}{\displaystyle m_\rt \, \rd m_\rt \, \rd y}$ 
distributions should be published. The conventional form of scientific 
publications may utilize long tables published in refereed journals to 
achieve such a goal. However, we would like to provide an additional method 
to reach the same purpose by the creation of the HBT and Spectrum database, 
to which the experimental collaborations can submit (or upload) data. The 
possibilities of electronic publishing can be exploited by the usage of 
such a database. 

\begin{figure}
\vspace{-0.5truein}
          \begin{center}
%          \leavevmode\epsfysize=7.0in
          \leavevmode\epsfysize=6.0in
          \epsfbox{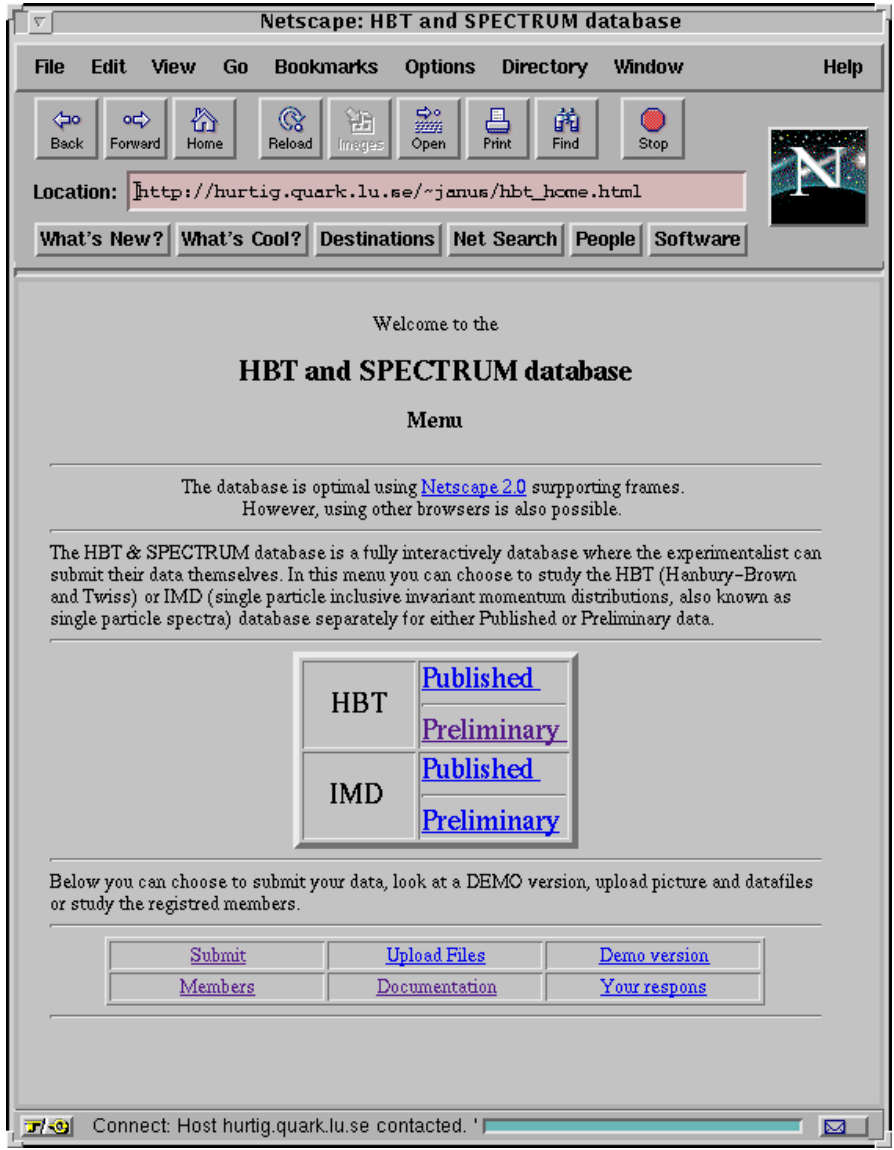}
          \end{center}
\caption{
The picture shows how the HBT and Spectrum database appears on the Internet 
when a Netscape browser is utilized HBT (Hanbury-Brown -- Twiss) stands for 
correlation functions, while IMD denotes single-particle inclusive invariant 
momentum distributions or particle spectra. The location of the page 
(WWW URL) is visible in the upper part of the picture: 
{\tt http://www.quark.lu.se/$\tilde{\,\,\,}$janus/hbt\underline{ }home.html}.
Note that 
{\tt www.quark.lu.se} and {\tt hurtig.quark.lu.se} refer to the same address.}
\label{fig:1}
\end{figure}

Thus, the HBT and Spectrum database provides a framework for the publication 
of one- or multi-dimensional distributions, in order to make the measured 
data points accessible electronically for the interested parties in an 
easy-to-use manner. Submitters to the database should be registered and 
should provide their e-mail address before they can get an automatically 
generated password which will be needed for submitting data to the database. 
(Of course, no password is requested from those who would like to download 
the data). Upload of files is also possible to provide pictures of the data 
and to make full distributions available for the interested scientific 
community. Some data are already available in a demo version of the database 
 - we suggest that you consult this demo before actually starting the 
transmission. A brief on-line documentation is also provided and the list of 
members or submitters to the database can be seen interactively.

\section{Responsibilities}

In order to give experimentalists full responsibility for their data, they
are supposed to submit the data themselves. This can be done by making a 
fully interactive database. Also, published and preliminary data should be 
clearly separated.
Published data are supposed to be fully corrected for known experimental
errors, the measured values and their statistical and systematic errors
are determined by the collaborations and published in refereed scientific
journals. On the other hand, sometimes it takes years until all the 
corrections and errors of the measurements are understood, and some data
are presented at conferences as the Quark Matter conference series~\cite{qm96}
 in a preliminary form. Such data are very useful to guide the development of
theoretical descriptions. However, they should not be confused with fully
corrected published data. In order to avoid such a case, the database
clearly distinguishes preliminary and published data, which could be submitted
and retrieved separately both for HBT and IMD data.  Submitters and users of
preliminary data are supposed to agree on the terms of the usage of preliminary
data. After some discussions with our experimental colleagues, the following
Agreement Form has been prepared:
{\it
``These data are supposed to be presented to the scientific community at some
conferences and/or other scientific meetings,  however they are not supposed 
to be final or fully corrected. Submission to the database of preliminary 
data implies that the submitter/experiment agrees to a broad distribution and 
discussion of their preliminary data, under the condition that the data are 
discussed as preliminary ones. Any user of the preliminary data is supposed 
to refer to these data as preliminary, i.e. not final data. The manager of 
the database is not responsible for any problems with the preliminary data, 
all responsibility to communicate correct research rests with the submitters 
and the users of the database."}

The database is fully interactive, which implies that the original submitter 
can edit or modify the submissions. The submitter can edit/delete the data 
submitted by him/herself at any time. A log file is created together with 
the submission, which is updated each time the submitted material is edited.

\section{Technical Description}

Utilization of a program language inside a HTML document (PHP/FI ver. 2.0) 
together with a database program (mSQL including a special interface W3-mSQL) 
made it possible to use the World-Wide-Web as interface between the user and 
the database. This ensures the  requested user-friendly and easily accessible 
environment. The preferable net-browser is Netscape, which  supports FRAMES 
and FORMS.  According to a recent estimate,  a Netscape  browser is utilized 
in more than 80 \% of Internet explorations. The browsers NCSA Mosaic or
Microsoft Explorer are also supported, but their usage results in
a less advanced performance.

Open the location 
{\tt http://www.quark.lu.se/$\tilde{\,\,\,}$janus/hbt\underline{ }home.html~}
in order to contact the database. Here one finds a menu where one can either 
have a look  at the database (presently separated into four parts, namely 
HBT and IMD for published and preliminary data, respectively) or submit data. 
In order to submit data one must register as a user. By doing this one will 
receive a password via e-mail. The submission includes: 
Name, title of paper, www-link to paper, author, submitted
when to whom, links to other papers and to the home page of the submitter,
 abstract, notes about the data, results in a table which can be
customized and kumac-, picture- and data-file upload (more about this below).
The submitted data  is sorted according to beam energy, beam particle and
type of particle measured.  

\section{Format for Data and Picture Files}
It is possible to upload ASCII and GIF files to this database. 
The ASCII datafile should preferably include as complete information
about the data as possible, for example the raw data set together with errors, 
various corrections, errors of corrections, and the fully corrected data 
set together with systematic and statistic errors. Such a policy is 
recommended in order to allow 
for new developments in the correction for the systematic errors on the data. 
For example, Coulomb corrections to measured two-particle correlation 
functions can be performed utilizing different methods,\cite{ztr,d96} and new 
methods for the Coulomb corrections can only be tested if the uncorrected 
data were available.

Below each data file further information should be included, like: colliding 
system, beam energy, acceptance, trigger condition, particle id definition, 
2 particle cuts/corrections, kinematic cuts, reference frame, fit function, 
fit parameters and errors, correlation contours for parameters and anything 
else that could be useful for the user to know.

\subsection{ Format of an {\mbox{ASCII}} data file  containing particle 
spectra}
It is very important for the use of the database, that also data points
for the fitted double-differential invariant momentum distributions 
and correlation functions will be available. 
This permits the performance of a more advanced analysis than
just to describe the fitted parameters of the data.

The format of the ASCII-file for the IMD or spectrum 
is indicated on Table~1. 

In Table 1, $p_i$, $i=1,2,3$ could be any momentum component, e.g. $p_x$,
$p_y$, $p_z$  in MeV,  or, alternatively, the rapidity and transverse mass, 
 $ y = 0.5 \log( (E+p_z)/(E-p_z)) $ and $m_\rt = \sqrt{ m^2 + p_x^2 + p_y^2}$, 
or a sub-set of the above.
Some comments appended to the end of the data file should specify  
the definitions and the  units of the  momentum variables. 
If not all the information
listed above is available, a fraction of it could be submitted, too. 

\newpage
\begin{center}
\vspace{0.5truecm}
{\bf Table 1}. Data structure for the single-particle spectrum\\[2mm]
%\end{center}
%\begin{center}
\begin{tabular}{|c c c c c |}
\hline\hline
$p_1$ & $p_2$ & $p_3$ & $N({\bf p})$ & $\delta N({\bf p})$\\ 
 .. &  .. &  .. &  .. &  ..   \\ 
 .. &  .. &  .. &  .. &  .. \\ 
\multicolumn{5}{|l|}{Comments: $p_1 = y  $, $ p_2 = m_t$ , ... }\\
\multicolumn{5}{|l|}{ Ref: Phys. Lett. .... } \\ \hline\hline
\end{tabular}
\vspace{0.5truecm}
\end{center}

From time to time, new expressions appear in the literature
which can describe the measured distributions in greater detail 
than the previously preferred formulae.
A well-known example for such a behavior is the out-long cross-term
in the two-particle correlation function,\cite{uli_s} another important
but not so well-known example is  
the appearance of an $m_\rt$ dependent effective
volume factor in the single-particle spectra,\cite{3d,halo}
which suggests that the particle spectra should be fitted with an expression
\begin{equation}
	N({\bf p}) = C \, m_\rt^{\alpha} \exp(-m_\rt / T_*(y)),
\end{equation}
where $\alpha$ is a new free fit parameter, the effective-slope parameter
$T_*$ should depend on the rapidity  $ y$ 
and $C$  stands for a  normalization constant. 
Previously, data were frequently fitted with the above expression utilizing
a fixed value of $\alpha = 0 $ or $\alpha = 0.5$ .
Because of this and similar reasons, we would like to encourage the
experimentalists to submit not only the fitted parameters but also 
 the measured distribution functions themselves to the database, in
as detailed a manner as possible.  

\subsection{ Format of ASCII File for Data on Correlations}

The following format is recommended for ASCII-files 
containing correlation functions:
\begin{center}
\vspace{0.5truecm}
{\bf Table 2}. Data structure for the  correlation functions
\vspace{0.2truecm}
\end{center}
\begin{center}
\small
\def\hh{\hskip-2.1mm}
\begin{tabular}{|ccccccccccc|} \hline\hline
$Q_1$ \hh & \hh $Q_2$ \hh & \hh $Q_3$ \hh & \hh A(Q) \hh & \hh  $\delta A(Q)$
 \hh & \hh  B(Q) \hh & \hh $\delta B(Q)$ \hh & \hh 
$C_{u}(Q) $\hh & \hh $\delta C_{u}(Q)$ \hh & \hh$C(Q)$\hh & \hh $\delta C(Q)$\\
.. \hh & \hh .. \hh & \hh .. \hh & \hh  .. \hh & \hh  .. \hh & \hh  .. \hh & 
\hh  .. \hh & \hh  .. \hh & \hh  .. \hh & \hh  .. \hh & \hh  .. \\ 
.. \hh & \hh .. \hh & \hh .. \hh & \hh  .. \hh & \hh  .. \hh & \hh  .. \hh & 
\hh  .. \hh & \hh  .. \hh & \hh  .. \hh & \hh  .. \hh & \hh  .. \\ 
\multicolumn{11}{|l|}{Comments: $Q_1=Q_{\inv}$, ...,  Ref: Phys. Rev... } \\ 
\hline\hline
\end{tabular}
\vspace{0.5truecm}
\end{center}
In Table 2, $Q = (Q_1,Q_2,Q_3)$ 
could be $Q=(Q_{\long}, Q_{\side}, Q_{\out})$ at a fixed value of $K$
for a 6-dimensional distribution, or $Q = Q_{\inv}$ for a
one dimensional distribution. (In the latter case the second and third
columns of the above table should of course be missing or left empty). 
 See e.g. refs.~\cite{bengt,zajc,urev} for the definitions of the components 
of the relative momentum. The relative-momentum components are to be 
specified in some comments appended to the end of the data file,
and a statement about the allowed region for the  mean momentum of the 
particles should be included with this comment, too. For a given value 
of the relative and mean momentum, we recommend submitting the Actual 
distribution of the particle pairs $A(Q)$, the Background 
distribution $B(Q)$, their errors $\delta A(Q)$ and $\delta B(Q)$, 
respectively, the uncorrected correlation function $C_{u}(Q) = A(Q)/B(Q)$, 
its error $\delta C_{u}(Q)$ and, finally, the fully corrected correlation 
function $C(Q)$ and its error $\delta C(Q)$. If the full information listed 
above is not  available in full details, a subset could be submitted, too. 

The ASCII data file together with the picture file of possible graphs should
be submitted utilizing the submit menu from 
the home page of the Correlations and Spectra database:\\
{\tt http://www.quark.lu.se/$\tilde{\,\,\,\,}$janus/hbt/scripts/upload.html}~.

\section{Future}

The development of the World-Wide-Web is fast and the possibilities are 
growing almost exponentially. The database will be updated regularly and 
suggestions for modifications or applying additional features will be taken 
seriously. The structure of the database is quite general, which allows for 
straight-forward extensions to include other type of data in case of future 
interest.

\section*{Acknowledgments:}
This database was prepared for the HBT'96: ``Particle Interferometry 
In High Energy Heavy Ion Physics" workshop, which took place at ECT*  in 
Trento, Italy, September 15 - 28, 1996. The work was reported first at 
the 7-th International Workshop on Multiparticle Production ``Correlations 
and Fluctuations", which was held in Nijmegen, The Netherlands, June 30 -- 
July 6, 1996. Cs. T. would like to thank W. Kittel for stimulating 
discussions, for a kind invitation to and support at the Nijmegen meeting. 
This work has been supported by the HNSF grants OTKA - F4019 and by an 
Advanced Research Award form the Fulbright Foundation, grant No. 20925/1996.

\end{document}